\documentclass[a4paper,english]{amsart}
\usepackage[T1]{fontenc}
\usepackage[latin1]{inputenc}
\usepackage{graphicx}
\usepackage{amssymb}

\makeatletter

 \theoremstyle{plain}    
 
 \numberwithin{equation}{section} 
 \numberwithin{figure}{section} 
 \theoremstyle{plain}    
 \newtheorem*{conjecture*}{Conjecture} 

\usepackage{babel}
\makeatother
\begin{document}
\def\CP{\mathbb{C}{\bf{P}}}
\bibliographystyle{/ada1/lpthe/viallet/formats/perso}

\title{On the complexity of some birational transformations.}


{\count255=\time\divide\count255 by 60 \xdef\hourmin{\number\count255}
	\multiply\count255 by-60\advance\count255 by\time
   \xdef\hourmin{\hourmin:\ifnum\count255<10 0\fi\the\count255}}

\def\oramin{\hourmin }

\def\ladate{\number\day/\ifcase\month\or janvier \or fevrier \or mars \or
avril \or mai \or juin \or juillet \or ao\^ut \or septembre
\or octobre \or novembre \or d\'ecembre \fi/\number\year\ (\oramin)}

\setbox200\hbox{$\scriptstyle \ladate $}

\def\lip{\copy200}


\author{J. Ch. Anglès d'Auriac$\dagger $, J. M. Maillard$\ddagger $ and
C. M. Viallet$\star $}
\address{$\dagger $ CRTBT-CNRS, BP 166, 38042 Grenoble FRANCE} 
\address{$\ddagger $ LPTL, 4 place Jussieu, 75252 Paris Cedex05 FRANCE}
\address{$\star $ LPTHE, 4 place Jussieu, 75252 Paris Cedex05 FRANCE.}

\date{\lip} 

\begin{abstract}
Using three different approaches, we analyze the complexity of various
birational maps constructed from simple operations (inversions) on
square matrices of arbitrary size.  The first approach consists in the
study of the images of lines, and relies mainly on univariate
polynomial algebra, the second approach is a singularity analysis, and
the third method is more numerical, using integer arithmetics.  Each
method has its own domain of application, but they give corroborating
results, and lead us to a conjecture on the complexity of a class of
maps constructed from matrix inversions.
\end{abstract}
\thanks{Work supported by Centre National de la Recherche Scientifique}
\maketitle

\begin{verbatim}


\end{verbatim}

\section{Presentation}
 \label{sec:Presentation}

The investigation of birational representations of Coxeter groups
acting on projective spaces of various dimensions appeared some years
ago to be of interest to understand the structure of lattice models of
statistical mechanics~\cite{BeMaVi91a,BeMaVi91c}.  Birational
dynamical systems have also been studied for their own sake with
various methods ranging from analysis to algebra.  A common ingredient
to these subjects is the study of iterations of infinite order
birational transformations, and in particular their complexity,
measured by the rate of growth of the degree of their iterates; see
for example
\cite{FaVi93,AbAnBoHaMa99,AbAnBoHaMa99b,RuSh97,BeVi99,BoFo00,Ta01,Ta01c,DiFa01,CaFa02,BeKi04}.

We perform this analysis for a definite class of transformations,
defined from elementary operations on matrices of size $q\times q$,
the entries of the matrices being the natural coordinates of complex
projective spaces $\CP_n$. Depending on the specific form of
the matrices, the dimension $n$ will take different values $(n\leq
q^2-1)$.

We explain, exemplify, and confront three different approaches to the
problem. We also present a conjecture for the value of the complexity
for a family of transformations of interest to statistical mechanics.

The paper is organized as follows. We state in section
\ref{sec:The-problem.}  the problem of calculating the complexity of a
birational transformation acting on a projective space, and define the
basic objects of interest. We introduce four families of maps, which
will be used for explicit calculations.  In section
\ref{sec:Computing-the-generating} we indicate how to infer the
generating function of the sequence of degree of iterates of a map
from its first terms. This provides a first method of calculation of
the complexity.  In section \ref{sec:Analytical-results:-the}, we
calculate exactly the sequence of degrees by an analysis of the
singularity structure for one of the families of maps.  In section
\ref{sec:Results.}, we describe an arithmetic approach, where we
examine the action of iterates on rational points (integer homogeneous
coordinates), and simply measure the growth of the size of the
coordinates. This yields approximate values of the complexity.  We
conclude with  a conjecture.

\section{The problem\label{sec:The-problem.}}

Let $K$ be a birational transformation of complex projective
space $\CP_n$. If we write $K$ in terms of homogeneous
coordinates, it appears as a polynomial transformation given by $n+1$
homogeneous polynomials of the same degree $d$. With the rule that we
should factorize out any common factor, $d$ is well defined in a given
system of coordinates. Of course it is {\em not} invariant by changes
of coordinates. We may construct the sequence $\{d_n\}$, of the
degrees of the iterates $K^{n}$ of $K$.

The growth of the sequence $d_n$ is a measure of the complexity of
$K$. In the absence of factorizations of the polynomials the
sequence would just be
\begin{eqnarray}
d_n=d_1^n=d^n.
\end{eqnarray}

What happens is that if some factorizations appear, they induce a drop
of the degree, so that we only have an upper bound
\begin{eqnarray}
\label{eq:bounddeg}
d_n \leq d^n.
\end{eqnarray}

The drop may even be so important that the growth of $d_n$ becomes
polynomial and not exponential anymore. A measure of the growth is the
algebraic entropy
\begin{eqnarray}
\epsilon = \lim_{n\rightarrow \infty} {1\over{n}} \log{ d_n},
\end{eqnarray}
or the complexity
\begin{eqnarray}
\lambda= \exp{ (\epsilon)}.
\end{eqnarray}

Both the entropy $\epsilon$ and the complexity $\lambda$ are invariant
by any birational change of coordinates. They are canonically
associated to the map $K$. Our aim is to calculate them for
definite classes of maps, which we now describe.

Suppose $M$ is a $q\times q$ matrix, and consider the two simple 
rational involutions $I$ and $J$: the involution $I$ is the matrix
inverse up to a factor (i.e. when written polynomially it amounts to
replacing each entry by its cofactor). The involution $J$ is the
element by element inverse (also called Hadamard inverse, which
replaces each entry $M_{ij}$ by its inverse $1/M_{ij}$).
The two involutions $I$  and $J$ do not commute, and their composition 
 $K=I\circ J$ is generically of infinite order.

The map $K$ acts naturally on ${\CP}_{q^2-1}$.  It is however possible
to define various reductions to smaller projective spaces in the
following way\cite{BeMaVi91e}. For a given size of square matrices, we
define a pattern as a set of equalities between entries of the
matrix. The set of all pattern is the set of all partitions of the
entries of the matrix. An example of a pattern is ``all diagonal
entries equal, all off-diagonal entries equal''. This corresponds to
the partition of the entries in two parts (diagonal +
off-diagonal). Clearly any pattern is preserved by the action of $J$.
We call admissible a pattern which is also stable by $I$ (or
equivalently $K$).

All admissible patterns have been classified for $q=4$ and some of
them for $q=5$ in~\cite{BeMaVi91e,BeMaVi91b,AnMaVi02b}.  It has been
also shown that $\lambda$ can vary considerably from one admissible
pattern to another.  For example for $5 \times 5$ cyclic {\em and}
symmetric matrices one has $\lambda =1$ (polynomial growth), whereas
with the cyclic matrices one gets $\lambda =(7+3\sqrt{5})/2$.

We will focus on four fundamental admissible patterns, which exist
whatever the size $q\times q$ of the matrix is. The first one is the
pattern $(S)$ of symmetric matrices. The second one $(C)$ is the
pattern of the cyclic matrices defined by $M_{i,j}=M_{i+1,j+1}$ (with
indices taken modulo $q$).  The third one is the pattern of matrices
which are at the same time cyclic and symmetric $(CS)$. The last one
is the general pattern $(G)$, without equality conditions between the
entries.

\section{A first approach:  generating functions}
\label{sec:Computing-the-generating}

From the sequence of degrees $\{d_n\}$, it is possible to construct a generating function
\begin{equation}
f(u)=\sum _{n=0}^{\infty }d_{n}u^{n}\label{eq:defserie}.
\end{equation}

Since the degrees are bounded by (\ref{eq:bounddeg}), the series
(\ref{eq:defserie}) always has a non zero radius of convergence
$\rho$. Actually
\begin{eqnarray}
\rho={1\over{\lambda}},
\end{eqnarray}

The calculation method is the following: calculate explicitly the
first terms of the series, and try to infer the values of the
generating function. The method is sensible if the generating function is
rational.

The striking fact is that indeed the generating function $f(u)$
happens to be a rational fraction with integer coefficients in most
cases. The consequence is that a finite number of terms of the series
determine it completely. For reversible maps (i.e. when there exists a
similarity relation between the map and its inverse), we have not
found any counterexample to this rule. There are however
non-reversible maps for which the generating function is not
rational~\cite{Pr05}. Another consequence of the rationality of $f$ is
that $\lambda$ is an algebraic integer, and we have no counterexample
yet to that.

For practical purposes, it is necessary to push the calculation of the
degree of the iterates as far as possible. Instead of evaluating the
full iterate, it is sufficient to consider the image of a generic line
$l$ with running point
\begin{eqnarray}
\label{genline}
l(t) = [ a_0+b_0\;t, a_1+b_1\; t, \dots, a_n+b_n \; t],
\end{eqnarray}
where $a_i, b_i$ are arbitrary coefficients, and evaluate the images
of $l(t)$ by $K^n$. The degree $d_n$ is read off from this image. The
calculation may furthermore be improved by using integer coefficients
in (\ref{genline}) and calculating (formal calculation software are
quite efficient at that) over polynomial with coefficients in
${\mathbb{Z}/\mathbb{Z}_p}$ with $p$ a sufficiently large prime
integer. Taking different values of $p$ and of the coefficients $a_i,
b_i$ helps eliminating the accidental simplifications which may occur.

Suppose we have the degree $d_{n}$ for the first values of $n$, say
$n=1\dots n_{\textrm{max}}$. We may fit the series with a Pad\'e
approximant $F$, with numerator (resp. denominator) of degree $N$ (
resp. $M$), such that
\begin{eqnarray}
N+M= n_{\textrm{max}}-1
\end{eqnarray}
$N$ running from $0$ to $ n_{\textrm{max}}-1$.  Our experience is
that, if $n_{\textrm{max}}$ is large enough, the rational fraction $F$
we find simplifies drastically, and stabilizes for some central values
of $N$.  This usually means that the exact generating function has
been reached.

Note that the expansion of the non optimal $\left[N,M\right]$ Padé
approximants yield non integer, or negative coefficients in the
expansion of $F$, in contradiction with these coefficients being a
degree. Table \ref{cap:Generating-functions-for} displays the
``exact'' expression we have inferred for the generating function for
various values of $q$ for the $(CS)$ pattern, as well as the value of
$m=N+M$ and the value of $n_{\textrm{max}}$.

When $n_{\textrm{max }}$ is larger than $m$, we have a prediction on
the next values of the degree, and this gives confidence that the
result is exact.

In Table \ref{cap:Generating-functions-for}, we also
give the inverse of the modulus of the smallest zero of the
denominator, as well as a numerical value computed as explained in
section~\ref{sec:Results.}.

\begin{table}
\begin{center}\begin{tabular}{|c|c|c|c|c|c|}
\hline 
$q$&
$f_{q}(u)$&
$n_{\textrm{max}}$&
$m$&
$\lambda $&
$\lambda _{\textrm{num}}$\\
\hline
\hline 
4$\left(\star \right)$&
\begin{tabular}{c}
\\
{\LARGE $\frac{\left(1+u\right)^{2}}{\left(1-u\right)^{2}}$}\\
\\
\end{tabular}&
$\infty $&
4&
1&
\\
\hline 
5$\left(\star \right)$&
\begin{tabular}{c}
\\
{\LARGE $\frac{\left(1+u+2u^{2}\right)^{2}}
{\left(1-u\right)^{3}\left(1+u+u^{2}\right)}$}\\
\\
\end{tabular}&
14&
9&
1&
1.0062\\
\hline 
6&
\begin{tabular}{c}
\\
{\LARGE $\frac{\left(1+2u\right)^{2}}{\left(1-u\right)\left(1-4u\right)}$}\\
\\
\end{tabular}&
15&
4&
4&
4.0003 \\
\hline 
7$\left(\star \right)$&
\begin{tabular}{c}
\\
{\LARGE $\frac{\left(1+u+3u^{2}\right)^{2}}
{\left(1-u\right)\left(1+u+u^{2}\right)\left(1-7u+u^{2}\right)}$}\\
\\
\end{tabular}&
12&
9&
6.854102&
6.8541\\
\hline 
8&
\begin{tabular}{c}
\\
{\LARGE $\frac{\left(1+u\right)\left(1+2u-u^{2}\right)}
{\left(1-u\right)\left(1-11u+7u^{2}-u^{3}\right)}$}\\
\\
\end{tabular}&
11&
7&
 10.331852&
10.3317\\
\hline 
9&
\begin{tabular}{c}
\\
{\LARGE $\frac{\left(1+u+3u^{2}-3u^{3}\right)^{2}}
{\left(1-u\right)\left(1-13u+2u^{2}+u^{3}
+12u^{4}-8u^{5}+u^{6}\right)}$}\\
\\
\end{tabular}&
11&
13&
12.832689&
12.8326\\
\hline 
10&
\begin{tabular}{c}
\\
{\LARGE $\frac{\left(1+3u\right)^{2}}
{\left(1-u\right)\left(1-18u+u^{2}\right)}$}\\
\\
\end{tabular}&
9&
5&
 17.944273&
17.9453 \\
\hline
11$\left(\star \right)$&
\begin{tabular}{c}
\\
{\LARGE $\frac{\left(1+u+5u^{2}\right)^{2}}
{\left(1-u\right)\left(1+u+u^{2}\right)\left(1-23u+u^{2}\right)}$}\\
\\
\end{tabular}&
7&
9&
22.956439&
22.9562\\
\hline
12&
\begin{tabular}{c}
\\
{\LARGE $\frac{\left(1+4u-3u^{2}\right)\left(1+2u-u^{2}\right)}
{\left(1-u\right)\left(1-27u+31u^{2}-9u^{3}\right)}$}\\
\\
\end{tabular}&
8&
8&
25.812541&
25.8105 \\
\hline
13$\left(\star \right)$&
\begin{tabular}{c}
\\
{\LARGE $\frac{\left(1+u+6u^{2}\right)^{2}}
{\left(1-u\right)\left(1+u+u^{2}\right)\left(1-34u+u^{2}\right)}$}\\
\\
\end{tabular}&
&
9&
33.970562&
33.9719\\
\hline
\end{tabular}\end{center}
\bigskip
\caption{\label{cap:Generating-functions-for}Generating functions for
the cyclic symmetric $(CS)$ patterns. The formulae for prime values of
$q$, tagged with a $\left(\star \right)$, can be proved.  $n_{\rm
max}$ is the maximum number of iteration performed, $m$ refers to the
Padé approximation, $\lambda$ is the complexity, $\lambda_{\rm num}$
is the numerical complexity calculated in section \ref{sec:Results.}}
\end{table}

\section{A second approach: Singularity analysis
\label{sec:Analytical-results:-the}}

In this section we prove that the complexity of the patterns $(CS)$
for prime $q$ is a quadratic integer, by showing that the sequence of
degrees verifies a linear recurrence relation of length 2 with integer
coefficients. This implies that the generating function of the
degrees is a rational fraction and corroborates a part of the results
given in table \ref{cap:Generating-functions-for}.

\subsection{Some notations}
\label{prelem}
Let $M$ be a cyclic symmetric matrix of size $q\times q$. The matrix
$M$ may be written in terms of the basic cycle of order $q$:\[
\sigma =\left(\begin{array}{ccccc}
 0 & 1 & 0 & \cdots  & 0\\
 0 & 0 & 1 & \cdots  & 0\\
 \cdots  & \cdots  & \cdots  & \cdots  & \cdots \\
 0 & 0 & 0 & \cdots  & 1\\
 1 & 0 & 0 & \cdots  & 0\end{array}
\right)\]
\[
M_{\textrm{even}}=x_{0}+x_{1}\left(\sigma +\sigma ^{q-1}\right)+
\cdots +x_{p-1} \; \sigma^{q/2},\qquad p=\frac{q}{2}+1\]
\[
M_{\textrm{odd}}=x_{0}+x_{1}\left(\sigma +\sigma ^{q-1}\right)+
\cdots +x_{p-1}\left(\sigma ^{(q-1)/2}+\sigma ^{(q+1)/2}\right),\qquad p
=\frac{q+1}{2}\]
when $q$ is even and odd respectively.

The parameter space is a projective space
${\mathbb{CP}_{p-1}}$ of dimension $p-1$, with $ p= q/2 +1$ if $q$ is
even and $ p= (q+1)/2$ if $q$ is odd. We use homogeneous coordinates
$[x_0, \dots, x_{p-1}]$.

We will study the two elementary transformations $I$ and $J$ acting on
$M$. Both are rational involutions (and are thus birational
transformations).

The Hadamard inverse $J$ may be written polynomially in terms of the
homogeneous coordinates
\begin{eqnarray} J : [x_0, \dots, x_{p-1}] \longrightarrow 
[\prod_{k\neq 0} x_k, \prod_{k\neq 1} x_k, \dots,  \prod_{k\neq p-1} x_k] 
\label{defj} 
\end{eqnarray} 
The matrix inverse $I$, up to a factor, transforms cyclic matrices
into cyclic matrices, and symmetric matrices into symmetric matrices.
It thus acts on cyclic symmetric matrices.

For cyclic symmetric matrices, the matrix inverse $I$ is related to
the Hadamard inverse $J$, by a similarity transformation:
\begin{eqnarray} 
\label{similarij}
 I = C^{-1} \circ J \circ C 
\end{eqnarray} 
The transformation $C$ acts linearly on the $p$ homogeneous
coordinates.  Denoting $\omega$ the $q$-th root of unity, $C$ is given
by the $p \times p$ matrix with entries:
\begin{eqnarray} 
C_{r,0}= 1, \quad C_{r,s}= \omega^{rs} + 
{\frac{1}{\omega^{rs}}} \quad \quad  r \neq 0 
\end{eqnarray} 
for $\, q $ odd and 
\begin{eqnarray} 
&&C_{r,0}= 1, \quad C_{r,s}= \omega^{rs} + 
{\frac{1}{\omega^{rs}}} \quad \quad  r \neq\,  0,\,  p-1,  \\
&& C_{r,p-1}= (-1)^r\,  \nonumber
\end{eqnarray} 
for $\, q $ even. 

The matrix $C$ verifies  $\, C^2 \, = {\bf 1}$


\subsection{Sequences of surfaces, sequences of degrees}

Consider now a sequence of hypersurfaces in ${\mathbb{CP}_{p-1}}$, 
obtained by applying successively $I$, then $J$, then $I$, and so on, 
starting with a generic hyperplane $S_0$. Each surface $S_n$ has a
polynomial equation, of degree $d_n$, which we also denote $S_n$. 
Since for non singular points,  
\begin{eqnarray} 
x \in S_{2n}  \quad\leftrightarrow  \quad J(x) \in S_{2n-1},  
\end{eqnarray} 
$S_{2n}$ can be obtained from $S_{2n-1}$ by substituting the
coordinates of $x$ with the homogeneous polynomial expression of the
coordinates of $J(x)$ in $S_{2n-1}(x)$.  Notice that, since $J$ is an
involution, $S_{2n-1}$ may be obtained from $S_{2n}$ in the same
manner.
\begin{figure}
\begin{center}\includegraphics[  scale=0.5]{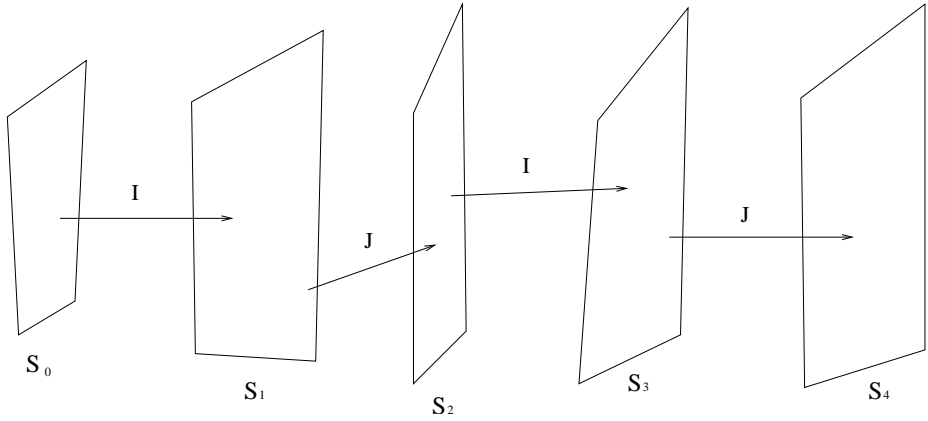}\end{center}
\caption{successive images}
\end{figure}

What happens at the level of the equations is that $S_{2n-1}(J(x))$
may factorize. One of the factors is $S_{2n}(x)$.  The only other
possible factors are powers of the coordinates of $x=(x_0, \cdots,
x_{p-1})$ as explained in the lemma below.  Relation
\begin{eqnarray}
\label{e1}  
S_{2n-1}(J(x)) = S_{2n}(x) \cdot \prod_{k=0}^{p-1}
x_k^{\alpha^{(k)}_{2n-1}}
\end{eqnarray} 
defines the exponents  $\alpha^{(k)}_{2n-1}$.

\subsection{A lemma\label{sub:A-lemma}}

The previous relation is crucial. Its proof is elementary and goes
as follows. 

Suppose $B$ is a birational involution.  When written in terms of the
homogeneous coordinates, $B^2$ appears as the multiplication by some
common polynomial factor of all the coordinates, that is to say the
identity transformation in projective space.
\begin{eqnarray} 
B (B (x)) = [\kappa_B (x) \cdot x_0,\kappa_B (x)  
\cdot x_2, \dots, \kappa_B (x) \cdot x_{p-1}] 
\end{eqnarray} 
with $\kappa_B (x) =$ some polynomial.

We then have, if two algebraic hypersurfaces
 $S$ and $S'$ are the proper images of 
each other by involution $B$: 
\begin{eqnarray} 
\label{ssigma} 
S(B(x)) &=& S'(x)\cdot R(x) \\
\label{spsi} 
S'(B(x)) &=& S(x) \cdot T(x) 
\end{eqnarray}
with $R $ and $T$ some polynomial expression of the coordinates.  
We then have 
using~(\ref{ssigma}) and~(\ref{spsi}): 
\begin{eqnarray} 
\kappa_B(x)^{{\rm deg}(S)} \cdot S(x) =   
S(x) \cdot  R(B(x)) \cdot T(x) 
\end{eqnarray} 
that is to say 
\begin{eqnarray} 
\label{cestla} 
\kappa_B(x)^{{\rm deg}(S)} =   R(B(x)) \cdot T(x) 
\end{eqnarray} 
Equation~(\ref{cestla}) shows that the only factors in the right 
hand side of equation~(\ref{spsi}) are the equation of $S$, 
and polynomial expressions $T(x)$ which divides $\kappa_B(x)$, 
possibly raised to some power. 

In the specific example $B=J$, and $S= S_{2\;n -1}$,   using  
\begin{eqnarray} 
\kappa_J(x) = \prod_{i=0}^{p-1} x_i^{p-2} \end{eqnarray} 
we get :
\begin{eqnarray} 
\label{Srho}
S_{2n-1}(J(x)) = S_{2n}(x) \cdot \prod_{i=0}^{p-1} x_i^{\rho_i},
\end{eqnarray} 
with $x_i(x)$ the $i$'th coordinate of $t$, and $\rho_i$ some
integer power. 

This ends the proof of formula~(\ref{e1}).

\subsection{Recurrence relation}
 Similarly to equation (\ref{e1}), we have: 
\begin{eqnarray} 
\label{e2} 
S_{2n}(J(x)) = S_{2n-1}(x)\cdot \prod_{i=0}^{p-1} x_i^{\alpha^{(i)}_{2n}},
\end{eqnarray} 
with the constraint
\begin{eqnarray} 
\label{kxx} 
\kappa_J(x)^{d_{2n}} =  \prod_{i=0}^{p-1} x_i^{\alpha^{(i)}_{2n-1}}(x)\cdot 
 \prod_{j=0}^{p-1} x_j^{\alpha^{(j)}_{2n}}(J(x)) .
\end{eqnarray}

We also have the corresponding equations for the action of $I$.
\begin{eqnarray}
\label{ee1}  
S_{2n}(I(x)) = S_{2n+1}(x) \cdot \prod_{k=0}^{p-1}
X_k^{\beta^{(k)}_{2n}}
\end{eqnarray} \begin{eqnarray} 
\label{ee2} 
S_{2n+1}(I(x)) = S_{2n}(x)\cdot \prod_{i=0}^{p-1} X_i^{\beta^{(i)}_{2n+1}} 
\end{eqnarray}
where $X_i$ is the i-th coordinate of $Cx$.

To make relations more uniform, we introduce a slight change of
notation: define the sequences $\{u_n^i\}$ and $\{v_n^i\}$ with the
convention that
\begin{eqnarray}
\alpha_{2k+1}^i = u_{2k+1}^i, \qquad \alpha_{2 k}^i = v_{2 k}^i, \\
\beta_{2k+1}^i = v_{2k+1}^i, \qquad \beta_{2 k}^i = u_{2 k}^i .
\end{eqnarray}

At  step $n$ we have $2p+1$ variables ($d_n$, $u_n^i$ and $v_n^i$).

A first equation simply expresses the factorization:
\begin{eqnarray}
d_n = \, (p-1)\, d_{n-1}-\sum_{i=0}^{p-1}u_{n-1}^i.
\label{eqdef}
\end{eqnarray}
Another set of equations is obtained by  expressing that both $I$
and $J$ are involutions:
\begin{eqnarray}
(p-2)\, d_n =\, v_{n-1}^i + \sum_{j \ne i} u_{n}^i, \qquad \quad i=0
\dots p-1.
\label{eqinv}
\end{eqnarray}
It is easy to get from equations (\ref{eqdef}) and
\ref{eqinv}:
\begin{eqnarray}
v_n^i &=& (p-2)\, d_{n-1}+ u_{n-1}^i  - \sum_{j=0}^{p-1} u_{n-1}^j, \,  
\quad \quad i=0 \dots p-1.
\label{sysdefinv2}
\end{eqnarray}

\subsection{Singularity structure}
\label{subsec:singstruct}

We need $p$ additional equations to complete the previous system. They
will be given, under some constraints, by the analysis of the
singularity structure. The basic idea is that the numbers
$\alpha_n^i,\; \beta_n^i$ (or equivalently $u_n^i, \; v_n^i$) have a
geometrical meaning: they are the multiplicity of some specific points
on the surface $S_n$.

The singularity structure of $J$ is very simple. A singular point is a
point whose image is undetermined: this happens when all polynomial
expressions giving the image~(\ref{defj}) vanish simultaneously.  Any
point with more than two vanishing coordinates is singular for $J$.

We will look at the action of the pair $I,J$ on the hypersurfaces
composing the factor $\kappa_J$ of eq. (\ref{kxx}). Those are
just the $n$ hyperplanes $\Pi_k$, $ k=0 \dots p-1$ of equation
\begin{eqnarray} 
\Pi_k : \qquad\{ x_k = 0 \}.
\label{Pik}
\end{eqnarray}
All intersections of these hyperplanes are made out of singular points
of $J$.  Some points are in a sense {\em maximally } singular. They
are the intersections of all but one of the planes $\Pi_r$, i.e. all
but one of their coordinates vanish.  There are $p$ such points
\begin{eqnarray} 
P_k = [ 0, \dots, 0,1,0, \dots, 0], \quad \quad \quad  
k\, =\, 0\, \dots\,  p-1 
\label{Pk}
\end{eqnarray} 
with $1$ in $(k+1)$-th position.

The singularity structure of $I$ is the same as the one of $J$, up to
the linear change of coordinates $C$. There are in particular $p$
distinguished singular points $Q_k$, $\, k=1 \dots p$ of $I$:
\begin{eqnarray} 
Q_k =\, C^{-1} \; P_k, \quad \quad k=0  \dots p-1. 
\end{eqnarray} 

To complete the set of equations (\ref{eqdef}), (\ref{eqinv}),
(\ref{sysdefinv2}) 
, we need to explore in some more details the
singularity structure of the maps.  What matters is the interplay
between $I$ and $J$.

The map $J$ sends the hyperplane $\Pi_k$ (\ref{Pik}) onto the point
$P_k$ (\ref{Pk}).  The subsequent images depend on what $q$ is.

The situation is tractable when $q$ is a prime number, in which case
the subsequent images of $\Pi_k$ {\em always go back to the point
$P_k$ after a finite number of steps}, actually one or three steps.
There, we meet a singularity, {\em and the equation of $\Pi_k$
factorizes}.  We will examine the case where $q$ is a prime number,
$q= 2p -1$.

The coordinate $x_0$ plays a special role and the point $P_0$ behaves
differently from the other points $P_s, \quad s=1 \dots p-1$.

Whatever $q$, the transformation of the hyperplane $\Pi_0$  reads :
\begin{eqnarray}  
\Pi_0 \mathop{\rightarrowtail}^{J} P_0 
\mathop{\longrightarrow}^I  P_0  
\mathop{\rightsquigarrow}^{J}\Pi_0 
\label{spp1} 
\end{eqnarray} 
We use the following convention concerning the arrows: when a variety
is sent by the map onto a variety of same codimension we use the plain
arrow $\longrightarrow$. When the codimension of the image is lower
(blow-down) we use the symbol $\rightarrowtail$, and when it is larger
(blow-up) we use the squiggly arrow $\rightsquigarrow$.  A blow-up for
the birational mapping $K$ corresponds to a blow-down for its inverse
$K^{-1}$.
 
The action of $I$ and $J$ on  the hyperplane $\Pi_s$ reads :
\begin{eqnarray} 
\Pi_s \mathop{\rightarrowtail}^{J} P_s 
\mathop{\longrightarrow}^I R_s 
\mathop{\longrightarrow}^J R_s
\mathop{\longrightarrow}^I P_s  
\mathop{\rightsquigarrow}^{J}\Pi_s 
\label{sppother} 
\end{eqnarray} 
The points $R_s$ have coordinates $[\pm 1, \pm 1, \dots, \pm 1]$. 
For example for $q=5$, $R_{2}=\left[+1,+1,-1\right]$ and 
$R_{3}=\left[+1,-1,+1\right]$,
while, for $q=7$,  $R_{2}=\left[+1,-1,-1,+1\right]$, 
$R_{3}=\left[-1,-1,+1,+1\right]$
and $R_{4}=\left[-1,+1,-1,+1\right]$. 

The pattern is similar for the points $Q_k$. It is obtained from the
previous one by the linear change of coordinates defined by $C$.  The
planes $ \Pi_k$ are replaced by the planes $ \Pi'_k = C^{-1} \Pi_k $
and the points $R_s$ are replaced by the points $R'_s = C^{-1}R_s$.

When $q$ is not a prime number, the pattern is different: the
successive actions of $I$ and $J$ leads to singular points other than
the $P_k$'s and $Q_k$'s. In appendix \ref{appendixq9} the case $q=9$ is
studied as an example.

Relations (\ref{spp1}), (\ref{sppother}) allow to relate the
multiplicities of the singular points $P_k$ on different surfaces
$S_n$.  Since $\,
P_0 \, \rightarrow \, P_0 \,$ in (Eq. \ref{spp1}) we have:
 \begin{eqnarray}
u_n^0  &=&  v_{n-1}^0
\label{ra1}
\end{eqnarray}
and since $\, P_s \, \rightarrow  \, P_s \,$ in (Eq. \ref{sppother}) 
we get: 
\begin{eqnarray}
 u^{s}_{n} &  =  &  v^{s}_{n-3}, \quad \quad s\,  =\,  1 \, \dots\,  p-1 
\label{ras}  
\end{eqnarray}

\subsection{End of the proof}
The previous analysis shows that when $q$ is prime, the factors $x_i$
(resp.  ${X}_i$) $1 \le i < p$ appear with the same exponent.  In
other words, for $q$ a prime number, the points $P_1$, $P_2$
... $P_{p-1}$ play an equivalent role, they will have the {\em same
multiplicities} on each $S_n$, and we will use $u_n^1$ to denote their
common value.

Using (\ref{ra1}), (\ref{ras}) together with (\ref{eqdef}) and
(\ref{sysdefinv2}) we get
\begin{eqnarray}
\label{recutil}
d_{n} & = & (p-1)\,d_{n-1}-u_{n-1}^{0}-(p-1)\,u_{n-1}^{1}, \nonumber \\
u_{n}^{0} & = & (p-2)\,d_{n-2}-(p-1)\,u_{n-2}^{1},\\
u_{n}^{1} & = & (p-2)\,d_{n-4}-u_{n-4}^{0}-(p-2)\,u_{n-4}^{1}.\nonumber
\end{eqnarray}
The rate of growth of the $d_n$'s is the inverse of the modulus of the
smallest eigenvalues of the $12 \times 12$ matrix given by the above
linear system. The outcome is  that the complexity of $K$ is the
inverse of the smaller root of
\begin{eqnarray}
\label{eq:compleC}
x^{2}+\left(2-\left(p-1\right)^{2}\right)x+1 = 0.
\end{eqnarray}

To get the full expression of the generating functions, we need to
specify the initial values of $d_n$, $u^0_n$ and $u^1_n$. They can
easily be calculated with the help of formal calculation software.
The results are summarized in table \ref{inival}.

\begin{table}
\begin{tabular}{|c|c|c|c|}
\hline 
$n$&
$d_{n}$&
$u_{n}^{0}$&
$u_{n}^{1}$\tabularnewline
\hline
\hline 
0&
1&
0&
0\tabularnewline
\hline 
1&
$p-1$&
0&
0\tabularnewline
\hline 
2&
$(p-1)^{2}$&
$p-2$&
0\tabularnewline
\hline 
3&
$p^{3}-3p^{2}+2p+1$&
$(p-1)(p-2)$&
0\tabularnewline
\hline
\hline 
4&
$(p-1)(p^{3}-3p^{2}+p+3)$&
$(p-1)^{2}(p-2)$&
$p-2$\tabularnewline
\hline
\end{tabular}
\caption{The initial values of $d_n$, $u^0_n$ and $u^1_n$ for $0 \le n
< 4$.  The values for $n=4$ have been deduced from the three previous
lines \label{inival}}
\end{table}

Note that when $\, q$ is not prime, we may still write a set of
recursions similar to (\ref{recutil}). The system is not complete, and
cannot be obtained from the analysis presented in
section~\ref{sec:Analytical-results:-the} (see Appendix
\ref{appendixq9}).

\section{Arithmetical approach: complexity through number of digits
\label{sec:Results.}}

The third approach consists in calculating the image of integer
points, and evaluating the growth of the size of the coordinates,
through the number of digits. It means that we do not try to calculate
the iterates formally.  This method was already experimented
in~\cite{AbAnBoHaMa99b}.

 Obviously the integer coordinates become extremely large, as large as
$10^{6000}$ and we used the library GMP to implement the
program\cite{gmp}. At each step of the calculation we factor out the
greatest common divisor of the components.  We assume that the
existence of a common factor between all the coordinates is due to a
factorization of the underlying polynomial. This assumption is valid,
at least after the first step where accidental factorization could
occur. The degree of the polynomial is estimated as the number of bits
used to store a typical entry (i.e. $\log _{2}(M_{ij})$ ).  The
algorithm proceeds as follows: i) construct a random matrix of
integers respecting the equalities of the pattern under consideration,
ii) replace each term by its cofactor, iii) divide every terms by the
greatest common factor of all of them, iv) replace each term by the
product of all others, v) divide every terms by the greatest common
factor of all of them, vi) record the number of digits used to store
the matrix elements. Note that one can exchange ii) and iv) without
altering the results. The procedure is iterated for as many steps as
possible, and possibly several runs with different initial matrices
are performed. Note that for pattern involving only very few variables
it can be efficient to write directly the recursion relation over the
variables.

The results are summarized in the Table \ref{cap:Algebraic-complexit},
giving  the value of the complexity for various
values of $q$ and for the four patterns introduced above. For cyclic
matrices and general $q$ it has been shown in ref~\cite{BeVi99} that
the value of the complexity $\lambda$ of $K = I \circ J $ is a
quadratic integer which is the inverse of the smaller root of
\begin{equation} 
x^{2}+\, (2- (q-2)^2) \, x +1. 
\end{equation}

In Table \ref{cap:Algebraic-complexit} an empty cell means that we
have not been able to compute the corresponding $\lambda$.  This is
due to the fast growth of the coordinates, preventing us to perform a
sufficient number of numerical iterations. The number of digits
displayed is just an indication of the estimated accuracy of our
numerical result.  When the values are known analytically we display
six digits.

\begin{table*}
\textbf{}\begin{tabular}{|c|c|c|c|c|}
\hline 
&
Cyclic Symmetric&
Cyclic&
Symmetric&
General\\
$q$&
$\lambda_{CS}$&
$\lambda_C$ &
$\lambda_S$ &
$\lambda_G$\\
\hline 
\hline 
$5$\begin{tabular}{c}
Numerical\\
Analytical\\
\end{tabular}&
\begin{tabular}{c}
1.00026\\
1\\
\end{tabular}&
\begin{tabular}{c}
6.85424 $\; \left[7\right]$\\
6.854102\\
\end{tabular}&
\begin{tabular}{c}
6.85972 $\; \left[7\right]$\\
\\
\end{tabular}&
\begin{tabular}{c}
6.85848 $\; \left[6\right]$\\
\\
\end{tabular}\\
\hline 
$6$\begin{tabular}{c}
Numerical\\
Analytical\\
\end{tabular}&
\begin{tabular}{c}
4.0003 $\; \left[10\right]$\\
4\\
\end{tabular}&
\begin{tabular}{c}
13.9288 $\; \left[5\right]$\\
13.928203\\
\end{tabular}&
\begin{tabular}{c}
13.8811 $\; \left[5\right]$\\
\\
\end{tabular}&
\begin{tabular}{c}
13.965 $\; \left[4\right]$\\
\\
\end{tabular}\\
\hline 
7\begin{tabular}{c}
Numerical\\
Analytical\\
\end{tabular}&
\begin{tabular}{c}
6.8541 $\; \left[7\right]$\\
6.854102\\
\end{tabular}&
\begin{tabular}{c}
22.9583 $\; \left[4\right]$\\
22.956439\\
\end{tabular}&
\begin{tabular}{c}
22.9771 $\; \left[4\right]$\\
\\
\end{tabular}&
\begin{tabular}{c}
22.972 $\; \left[4\right]$\\
\\
\end{tabular}\\
\hline 
$8$\begin{tabular}{c}
Numerical\\
Analytical\\
\end{tabular}&
\begin{tabular}{c}
10.3317 $\; \left[6\right]$\\
\\
\end{tabular}&
\begin{tabular}{c}
33.972 $\; \left[4\right]$\\
33.970562\\
\end{tabular}&
\begin{tabular}{c}
33.970 $\; \left[3\right]$\\
\\
\end{tabular}&
\begin{tabular}{c}
34.118 $\; \left[3\right]$\\
\\
\end{tabular}\\
\hline 
$9$\begin{tabular}{c}
Numerical\\
Analytical\\
\end{tabular}&
\begin{tabular}{c}
12.8326 $\; \left[5\right]$\\
\\
\end{tabular}&
\begin{tabular}{c}
47.027 $\; \left[3\right]$\\
46.978714\\
\end{tabular}&
\begin{tabular}{c}
47.040 $\; \left[3\right]$\\
\\
\end{tabular}&
\begin{tabular}{c}
47.000 $\; \left[3\right]$\\
\\
\end{tabular}\\
\hline 
$10$\begin{tabular}{c}
Numerical\\
Analytical\\
\end{tabular}&
\begin{tabular}{c}
17.9453 $\; \left[4\right]$\\
\\
\end{tabular}&
\begin{tabular}{c}
62.004 $\; \left[3\right]$\\
61.983868\\
\end{tabular}&
\begin{tabular}{c}
62.091$\; \left[3\right]$\\
\\
\end{tabular}&
\begin{tabular}{c}
62.085 $\; \left[2\right]$\\
\\
\end{tabular}\\
\hline 
$11$\begin{tabular}{c}
Numerical\\
Analytical\\
\end{tabular}&
\begin{tabular}{c}
22.9562 $\; \left[4\right]$\\
22.956439\\
\end{tabular}&
\begin{tabular}{c}
79.02 $\; \left[3\right]$\\
78.987340\\
\end{tabular}&
\begin{tabular}{c}
79.133 $\; \left[2\right]$\\
\\
\end{tabular}&
\begin{tabular}{c}
80.711 $\; \left[2\right]$\\
\\
\end{tabular}\\
\hline 
$12$\begin{tabular}{c}
Numerical\\
Analytical\\
\end{tabular}&
\begin{tabular}{c}
25.8105 $\; \left[4\right]$\\
\\
\end{tabular}&
\begin{tabular}{c}
98.03 $\; \left[3\right]$\\
97.989795\\
\end{tabular}&
\begin{tabular}{c}
99.17 $\; \left[2\right]$\\
\\
\end{tabular}&
\begin{tabular}{c}
100.32 $\; \left[2\right]$\\
\\
\end{tabular}\\
\hline 
$13$\begin{tabular}{c}
Numerical\\
Analytical\\
\end{tabular}&
\begin{tabular}{c}
33.972 $\; \left[3\right]$\\
33.970562\\
\end{tabular}&
\begin{tabular}{c}
130.3 $\; \left[3\right]$\\
118.9916\\
\end{tabular}&
\begin{tabular}{c}
121.6 $\; \left[2\right]$\\
\\
\end{tabular}&
\begin{tabular}{c}
121.5 $\; \left[1\right]$\\
\\
\end{tabular}\\
\hline 
$14$\begin{tabular}{c}
Numerical\\
Analytical\\
\end{tabular}&
\begin{tabular}{c}
39.169 $\; \left[2\right]$\\
\\
\end{tabular}&
\begin{tabular}{c}
142.8 $\; \left[2\right]$)\\
141.9930\\
\end{tabular}&
\begin{tabular}{c}
144.5 $\; \left[2\right]$\\
\\
\end{tabular}&
\begin{tabular}{c}
144.2 $\; \left[1\right]$\\
\\
\end{tabular}\\
\hline 
$15$\begin{tabular}{c}
Numerical\\
Analytical\\
\end{tabular}&
\begin{tabular}{c}
42.19 $\; \left[2\right]$\\
\\
\end{tabular}&
\begin{tabular}{c}
167. $\; \left[2\right]$\\
166.9940\\
\end{tabular}&
\begin{tabular}{c}
170. $\; \left[2\right]$\\
\\
\end{tabular}&
\begin{tabular}{c}
\\
\\
\end{tabular}\\
\hline 
$16$\begin{tabular}{c}
Numerical\\
Analytical\\
\end{tabular}&
\begin{tabular}{c}
49.10 $\; \left[2\right]$\\
\\
\end{tabular}&
\begin{tabular}{c}
194. $\left[2\right]$\\
193.9948\\
\end{tabular}&
\begin{tabular}{c}
\\
\\
\end{tabular}&
\begin{tabular}{c}
\\
\\
\end{tabular}\\
\hline 
$17$\begin{tabular}{c}
Numerical\\
Analytical\\
\end{tabular}&
\begin{tabular}{c}
61.66 $\; \left[2\right]$\\
61.983868\\
\end{tabular}&
\begin{tabular}{c}
224. $\left[2\right]$\\
222.9955\\
\end{tabular}&
\begin{tabular}{c}
\\
\\
\end{tabular}&
\begin{tabular}{c}
\\
\\
\end{tabular}\\
\hline
\end{tabular}
\bigskip

\caption{Complexities of $K = I \circ J $ for various values of $q$,
 for patterns $(CS)$, $(C)$, $(S)$ and $(G)$. The numerical and
 analytical results are displayed. The number in brackets is the
 number of iterations of $ K$ we have been able to
 calculate. \label{cap:Algebraic-complexit}}
\end{table*}

\section{Conclusion\label{sec:Conclusion.}}

The three different approaches we have used give corroborating
results. This gives us very good confidence both in the heuristic
method of section \ref{sec:Computing-the-generating}, and the more
numerical approach of section~\ref{sec:Results.}, thanks to the proof
given in section \ref{sec:Analytical-results:-the}.  We see by
comparing the two last columns of table \ref{cap:Algebraic-complexit}
 that $\lambda_G$ happens to be extremely close to $\lambda_S$,
as well as to $\lambda_C$. This allows us to state the conjecture:

\begin{conjecture*} 
The complexity of the transformation $K=I\circ J$ for the general
matrices (pattern $(G)$), for symmetric matrices (pattern $(S)$), and
for cyclic matrices (pattern $(C)$), are the same. Their common value
is the inverse of the modulus of the smaller root of $\quad x^{2} - ( q^2
-4 \; q+2)\;  x+1 = 0$.
\end{conjecture*}

Such a result would mean that although the number of parameters of
pattern (G) and (S) is much bigger than the one of pattern (C), the
latter captures the entirety of the complexity of the product of
inversions $K=I\circ J$. This might be related to the structure of
bialgebra of the set of square matrices equipped with ordinary matrix
product and Hadamard product. Phrased differently, the ``squeleton''
formed by the cyclic matrices encodes the structure of the whole
bialgebra.  This deserves further investigations which are beyond the
scope of this paper.

\appendix
\section{The cyclic symmetric case for  $q=9$}
\label{appendixq9}
We  consider in this appendix the case $q=9$.  Since $q$ is not a
prime number, our result of section \ref{sec:Analytical-results:-the}
does not apply.

The number of homogeneous variables is $p = (q+1)/2  =5$.  We use
the same notation as in the text for the hyperplane $\Pi_k$ and the
point $P_i$. In addition we define the three points
$Q_{1}=(1,1,-1,-1,1)$, $Q_{2}=(1,1,1,-1,-1)$ and
$Q_{4}=(1,-1,1,-1,1)$. We also introduce the codimension-two variety
$\Pi_{0,3}$ defined by the two equations $x_0=0$ and $x_3=0$. The
singularity structure is:
\begin{eqnarray*} 
& \Pi_0 &  \mathop{\rightarrowtail}^{J} P_0  \mathop{\longrightarrow}^I  P_0 
          \mathop{\rightsquigarrow}^{J}\Pi_0    \\ 
& \Pi_1 & \mathop{\rightarrowtail}^{J} P_1 \mathop{\longrightarrow}^I  Q_1 
           \mathop{\longrightarrow}^J Q_1 \mathop{\longrightarrow}^I P_1 
           \mathop{\rightsquigarrow}^J \Pi_1    \\ 
& \Pi_2 & \mathop{\rightarrowtail}^{J} P_2 \mathop{\longrightarrow}^I  Q_2 
           \mathop{\longrightarrow}^J Q_2 \mathop{\longrightarrow}^I P_2 
           \mathop{\rightsquigarrow}^J \Pi_2    \\ 
& \Pi_3 & \mathop{\rightarrowtail}^{J} P_3 \mathop{\longrightarrow}^I  Q_3 
           \mathop{\rightsquigarrow}^{J} \Pi_{0,3}    \\ 
& \Pi_4 & \mathop{\rightarrowtail}^{J} P_4 \mathop{\longrightarrow}^I  Q_4 
           \mathop{\longrightarrow}^J Q_4 \mathop{\longrightarrow}^I P_4 
           \mathop{\rightsquigarrow}^J \Pi_4 
\end{eqnarray*}
the subsequent iterates of $\Pi_{0,3}$ are non singular.  We see that
there will be six sets of exponents, $u^0_n$ and $v_n^0$ related to
$x_0$, $u^1_n$ and $v_n^1$ related to $x_1$, $x_2$ and $x_4$, and
finally $u^2_n$ and $v_n^2$ related to $x_3$.  The equations
expressing the degree drop due to the factorization, and the fact that
$I$ and $J$ are involutions, are:
\begin{eqnarray*}
d_{n+1}   & = & 4 d_n - u^0_n - 3 u^1_n - u^2_n,  \nonumber    \\
v^0_{n+1} & = & 3 d_n -         3 u^1_n - u^2_n,     \\
v^1_{n+1} & = & 3 d_n - u^0_n - 2 u^1_n - u^2_n,  \nonumber    \\
v^2_{n+1} & = & 3 d_n - u^0_n - 3 u^1_n.      \nonumber       
\end{eqnarray*}
Moreover the singularity structure shown above yield :
\begin{eqnarray*}
u^0_{n+1} & = & v^0_{n} \nonumber \\
u^1_{n+1} & = & v^1_{n-2}\nonumber 
\end{eqnarray*}
It is clear that an equation is missing to close the system:
\begin{eqnarray*}
\label{recu9}
d_{n+1}   & = & 4 d_n   - u^0_n - 3 u^1_n - u^2_n  \nonumber      \\
u^0_{n+1} & = & 3 d_{n-1} -         3 u^1_{n-1} - u^2_{n-1}     \\
u^1_{n+1} & = & 3 d_{n-3} - u^0_{n-3} - 2 u^1_{n-3} - u^2_{n-3} \nonumber    
\end{eqnarray*}
If we suppose that there exists a recursion relation of the form:
\begin{eqnarray*}
u^2_{n+1} \, = \, \, a \; d_{n-q} \, + b \; u^0_{n-q} \, 
+ c \; u^1_{n-q} \, +d  \; u^2_{n-q}\, + e ,
\end{eqnarray*}
where the $\, a, \, b, \, c, \, d, \, e$, as well as the shift $\, q$,
are integer constants. 
The hypothesis $\, q \, = \, 1$  yields:
\begin{eqnarray}
\label{missing}
u^2_{n+1} \, = \, \, 2 \, d_{n-1} \, -3 \,u^1_{n-1}  .
\end{eqnarray}
Introducing, with obvious notations, the generating functions 
\begin{eqnarray}
d(s) \, = \, \sum d_n \, s^n, \qquad u_i(s)\, = \, \sum u^i_{n} \,
s^n, \qquad i \, = \, 1, \, 2, \, 3 \nonumber
\end{eqnarray}

\begin{table}
\begin{center}
\begin{tabular}{|c|c|c|c|c|}
\hline 
$n$& $d_{n}$& $u_{n}^{0}$& $u_{n}^{1}$& $u_{n}^{2}$
\tabularnewline
\hline
\hline 
0& 1& 0& 0&
0\tabularnewline
\hline 
1& 4& 0& 0&
0\tabularnewline
\hline 
2& 16& 3& 0&
2\tabularnewline
\hline 
3& 59& 12& 0&
8\tabularnewline
\hline
\hline 
4& 216& 46& 3&
32\tabularnewline
\hline
\end{tabular} \end{center}
\bigskip \bigskip
\caption{The initial values of $d_n$, $u^0_n$, $u^1_n$ and $u^2_n$ for
 $0 \le n \le 4$.
 \label{inival2}}
\end{table}

one easily finds:
\begin{eqnarray*}
\label{finald9}
&& d(s) \, = \, \,
1\, +{\frac {(4 -s^2 -s^6)\, s }{P(s) }}, \qquad 
 u_0(s) \, = \, \,{\frac { ( 2\,{s}^{2}-3)\, (1+s^2) \,  s^4  }{P(s) }},
 \qquad \\
&& u_1(s) \, = \, \,{\frac { ( 2\,{s}^{2}-3) \,  s^4   }{P(s) }}, \qquad 
 u_2(s) \, = \, \,{\frac {(3\,{s}^{4}-2\,{s}^{2}-2)\, {s}^{2}  }{P(s) }}, 
\qquad 
\end{eqnarray*}
with :
\begin{eqnarray*}
P(s) \, = \, \, \left( 1-s \right) \cdot \left(
1-3\,s-2\,{s}^{2}-{s}^{3}+2\,{s}^{4}+2\,{s}^{5}-{s}^{6} \right)
\end{eqnarray*}
from which
\begin{eqnarray*}
f_9(u) =  \, \frac{
(1+u+3\, u^2\, -3\, u^{3})^2} {(1-u) \, (1\, -13\, u\, +2\, u^2+u^3
+12\, u^4\, -8\, u^5+u^6)} \nonumber
\end{eqnarray*}

\bibliography{/ada1/lpthe/viallet/papiers/ref}

\end{document}